\documentclass[11pt,aps,floats,amssymb,tightenlines,nofootinbib]{revtex4}
\usepackage{color}
\usepackage{graphicx}
\usepackage{amsmath}
\usepackage{amsfonts}
\usepackage{amscd}
\usepackage{epsfig}
\usepackage{amssymb}
\usepackage{tabularx}

\begin{document}

\title{Thermodynamic instability of splitting thin shell solutions in braneworld Einstein-Gauss-Bonnet gravity}
\author{Marcos Ramirez}
\date{December 2022}
\affiliation{IATE, CONICET - Universidad Nacional de C\'ordoba}

\begin{abstract}
   The thermodynamic stability of braneworld cosmological solutions in five-dimensional Einstein-Gauss-Bonnet gravity is addressed, particularly in the context of the splitting vacuum thin shells in which the shells represent either an ephemeral or a branch-changing false vacuum bubble spontaneously emerging from the braneworld and there is more than one possible solution for the same distributional initial data set. The free energy associated with the competing solutions is computed. It is shown that the criterion to decide which solution is preferred coincides with the criterion of the possibility of having a splitting solution. This result suggests an equivalence between a thermodynamic and a dynamic selection criteria among possible classical solutions that deserve further consideration.
\end{abstract}

%agregar keywords: las mismas que el paper anterior, más thermodynamic stability

\maketitle

\section{Introduction}

Lovelock gravity is arguably the most natural generalization of general relativity to arbitrary spacetime dimensions. It has several nice properties such as second order and quasi-linear field equations and that it reduces exactly to General Relativity (GR) when restricted to four dimensions. In five or six dimensions, Lovelock gravity reduces to Einstein-Gauss-Bonnet gravity (EGB), which is a theory whose field equations are of second order in curvature. However, this theory has several shortcomings. There are causality issues common to any higher-order in curvature gravity theory: gravitational perturbations might not follow light-like trajectories and there can be superluminal propagation of signals. In any case, in some situations (e.g. when curvature is sufficiently weak) causality can be recovered by defining the causal structure according to gravitational dynamics \cite{BrusteinSherf}, but this is not always possible. Moreover, at the moment there is no well-posed initial value formulation for these theories and there is evidence in the contrary \cite{ReallPapallo}. Notwithstanding these shortcomings, the analysis of solutions of Lovelock gravity continues to be an active field of research. In the early 2000's, there was a renewed interest in EGB theory because of the emergence of phenomenological braneworld gravity models. It was then natural to incorporate the EGB term when considering one large extra dimension, as Lovelock gravity is a natural candidate in order to take into account higher curvature corrections of the Einstein-Hilbert action coming from a low-energy limit of a quantum gravity theory. Suitable junction conditions for the matching of two spacetimes through a thin shell in EGB gravity where derived \cite{GravanisWillison1,Davis}. These conditions relate the jump of a third-order expression in the extrinsic curvature with the stress-energy tensor defined on the shell. They, being of third order, allow the matching of spacetimes in a non-trivial way that is impossible in pure GR: through a thin shell devoid of any matter-energy but with a discontinuous extrinsic curvature, the so-called {\it vacuum thin shell} \cite{GravanisWillison2,Garraffo}.    

The existence of vacuum thin shells allowed the exploration of a new kind of phase transition in semiclassical gravity, the {\it thermalon mediated} phase transitions \cite{Thermalon1,Thermalon2,Thermalon3}. Vacuum solutions of Lovelock gravity in spherical symmetry can be separated into different {\it branches}. There is a finite set of different solutions, with only one of them having as a limiting case the unique Schwarzschild-Tangherlini solution of GR (which we call the GR branch). It has been shown \cite{Garraffo} that, in a spherically symmetric setting, a vacuum thin shell may constitute a bubble (that is, may glue an interior with an exterior solution) only if it glues two different branches, that is, if it is a {\it vacuum bubble} in the sense of finite temperature Quantum Field Theory (QFT), that might be either {\it true} (if it encloses a GR branch solution) or {\it false} (if it encloses any other branch). In this way, a non-trivial dynamics involving vacuum solutions of Lovelock gravity has been discovered when taking into account the thermodynamic potentials associated with static vacuum thin shell solutions representing {\it vacuum bubbles} and their classical dynamics.

In a previous work by this author, a new family of solutions of EGB gravity involving vacuum thin shells and a braneworld was derived \cite{Ramirez1}. They represent the emergence out of a braneworld of a {\it false vacuum bubble} in the bulk, and they can be understood as braneworlds undergoing a splitting at a given point of their evolution. The vacuum bubble might be ephemeral (it might collide with and be reabsorbed by the braneworld) or it might collapse into the bulk black hole thereby changing the branch of the bulk solution. 
They have an interesting property already found in GR involving splitting thin shells \cite{Ramirez2}: whenever these solutions are possible, a non-splitting solution (the original evolving shell without any splitting) is also possible. This compromises the uniqueness of the evolution for a given initial data set. 

In this article, it is proposed to compute a suitable thermodynamic potential among competing solutions in order to determine whether one of them is semiclassically preferred. This involves the non-trivial task of computing Euclidean actions of dynamic solutions, addressed in Section \ref{energy}, for which we resort to approximations. It will be shown that the thermodynamically preferred solution is always the one without the splitting: there seems to be a non-trivial connection between the purely dynamical criteria that defines the possibility of the splitting and the semiclassical one that defines the thermodynamic preference. This apparent coincidence may signal a deeper connection between purely dynamical and semiclassic thermodynamical stability analysis.

% and now comment on how natural an emanating vacuum bubble is!!!!

\section{A vacuum bubble emanating from a braneworld}
\label{EGBsplitting}

Let us recall the EGB gravity action

\begin{equation}
I=\frac{1}{2\kappa^2}\int d^D x \sqrt{-g}(R-2\Lambda+\alpha L_{GB}) + I_m \;\; , \;\; L_{GB}=R^2-4R_{\mu\nu} R^{\mu\nu}+R_{\alpha\beta\mu\nu}R^{\alpha\beta\mu\nu} ,
\end{equation}

and the associated junction conditions through a thin shell \cite{Davis}

\begin{equation}
\label{junction}
[Q^a_b]_{\pm}=-\kappa^2 S^a_b \;\;\; , \;\;\; Q^a_b:=K^a_{b}-\delta^a_bK+2\alpha(3J^a_b-\delta^a_bJ-2P^a_{cbd}K^{cd}),
%[K^a_{b}]_{\pm}-\delta^a_b[K]_{\pm}+2\alpha(3[J^a_b]_{\pm}-\delta^a_b[J]_{\pm}-2P^a_{cbd}[K^{cd}]_{\pm})=-\kappa^2S^a_b,
\end{equation}
where latin indexes denote tensor quantities defined in the shell submanifold, $S^{ab}$ is the stress-energy tensor on the shell and 

\begin{eqnarray}
J_{ab}&:= &\frac{1}{3}(2KK_{ad}K^d_b + K_{df}K^{df}K_{ab} - 2K_{ad}K^{df}K_{fb}- K^2K_{ab}), \\
P_{adbf}&:= &R_{adbf} + 2h_{a[f}R_{b]d} + 2h_{d[b}R_{f]a} + Rh_{a[b}h_{f]d}.
\end{eqnarray}
These junction conditions are derived from a generalized Gibbons-Hawking boundary term

\begin{equation}
  {\cal I}_{\partial M} = - \frac{1}{\kappa^2} \int_{\partial M} K+2\alpha(J-2G^{ab}K_{ab}),   
\end{equation}
where $G_{ab}$ is the $D-1$-dimensional Einstein tensor for the metric on the boundary.

Let us consider an isotropic 5-dimensional spacetime in the sense that it is foliated by maximally symmetric $3$-surfaces. By virtue of a generalized Birkhoff's theorem \cite{Zegers} the metric has the form
\begin{equation}
\label{metric}
ds^2=-f(r)dt^2+f^{-1}(r)dr^2+r^2d\Sigma_k^2 \;\;\;\; , \;\;\;\; f(r)=k+\frac{r^2}{4\alpha}\left(1+\xi\sqrt{1+\frac{4\alpha\Lambda}{3}+\frac{\mu\alpha}{r^4}}\right),
\end{equation}
where $k=-1,0,1$, $\mu$ is an integration constant associated with mass and $\xi=\pm 1$ (each value representing a different branch). In order to have large $r$ asymptotics and the possibility to include maximally symmetric solutions one should impose $1+\frac{4}{3}\alpha\Lambda >0$. As usual in string inspired models we will also impose $\alpha>0$. 

Our system has a braneworld setting: a thin shell and a bulk that has $Z_2$-symmetry centered on the shell. The symmetries imposed imply that the intrinsic metric of the brane is a FLRW metric
\begin{equation}
ds^2_{\cal{S}}=-d\tau^2+a(\tau)^2d\Sigma_{k}^2 ,
\end{equation}
where $a(\tau)$ is the $r$ coordinate evaluated on the shell, and that the stress-energy tensor in the shell can be written as that of a comoving perfect fluid $S_a^b=diag[-\rho,p,p,p]$. As usual in braneworld settings, a brane tension $\sigma$ must be imposed in order to be able to recover standard Friedmann evolution at late times. 

The $\tau \tau$ component of the junction conditions (\ref{junction}) is written in this case as follows

\begin{equation}
\label{Q}
[Q^{\tau}_{\tau}]=\frac{6}{a^3}\sqrt{\dot{a}^2+f(a)}\left(a^2+4\alpha\left(k+\frac{2}{3}\dot{a}^2-\frac{1}{3}f(a)\right)\right) = \kappa^2(\rho + \sigma),
\end{equation}
from which an equation of motion can be obtained
\begin{equation}
\label{branemotion}
H^2+V_{\mu,\xi}(a)=0,
\end{equation}
where, as usual, $H=\dot{a}/a$. The effective potential can be written as
\begin{equation}
V_{\mu,\xi}(a)=-\frac{1}{8\alpha}\left[B_{\xi}(P(a)^2,A_{\mu}(a)^{3/2})+\frac{A_{\mu}(a)}{B_{\xi}(P(a)^2,A_{\mu}(a)^{3/2})}-2-\frac{8k\alpha}{a^2} \right],  
\end{equation}
where
\begin{equation}
B_{\xi}(P^2,A_{\mu}^{3/2})=-\xi A_{\mu}^{3/2}+256\alpha^3P^2+16\sqrt{2\alpha^3P^2(128\alpha^3P^2-\xi A_{\mu}^{3/2})} ,
\end{equation}
and the functions $P(a)$ and $A_{\mu}(a)$ are defined by
\begin{equation}
\label{AP}
P(a)=\frac{\kappa^2}{16\alpha}\left(
\rho+\sigma \right) \;\;\;\; , \;\;\;\; A_{\mu}(a)=\beta^2+\frac{\alpha\mu}{a^4}.
\end{equation}

\begin{figure}
\centerline{\includegraphics[width=.2\textwidth]{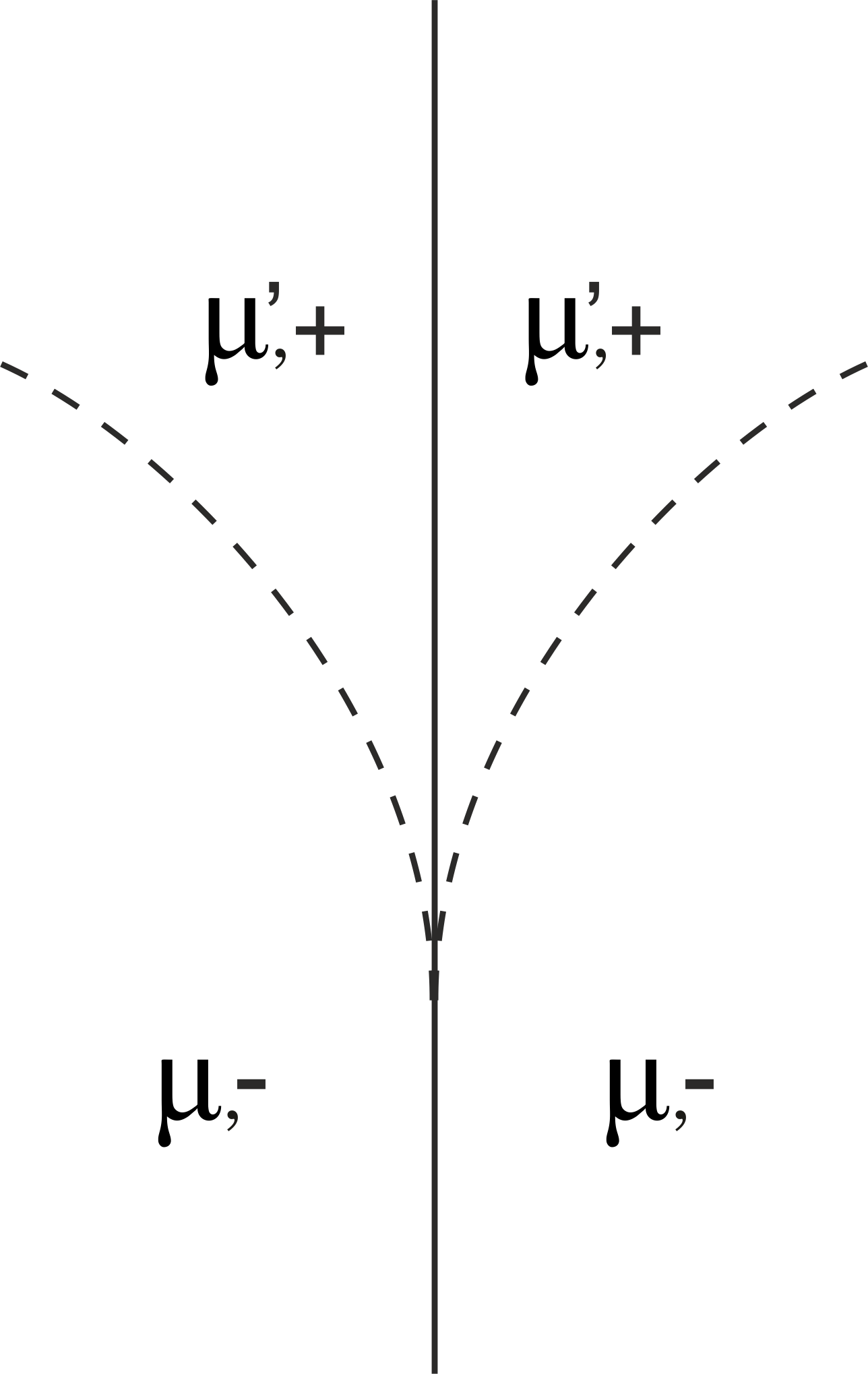}}
\caption{Illustration of the splitting solutions whose existence was proved in \cite{Ramirez1}}
\label{vacuumsplitting}
\end{figure}

As mentioned in the Introduction, this author proved in \cite{Ramirez1} the existence of solutions that represent a pair of vacuum thin shells (respecting $Z_2$ symmetry) emanating from the central braneworld, as illustrated in Fig \ref{vacuumsplitting}. The situation there considered is the spontaneous creation of a false vacuum bubble out of the central brane. The bulk spacetime changes in the vicinity of the brane from an initial GR branch solution with mass parameter $\mu$ to a stringy branch with mass parameter $\mu'$. The splitting is as smooth as possible in the sense that the tangent spaces of the different submanifolds at the separation surface coincide, and this condition sets $\mu'$ as a function of the initial parameters of the configurations. We now define
\begin{eqnarray}
x(a)=\frac{A^{3/2}_{\mu}(a)}{256\alpha^3P(a)^2} \;\; &,& \;\; y(a)=\frac{A^{3/2}_{\mu'}(a)}{256\alpha^3P(a)^2}  \\
\omega(x)=(1+x+\sqrt{1+2x})^{1/3} \;\;&,&\;\; \nu(y)=(1-y+\sqrt{1-2y})^{1/3}  \\
 g(x)=\omega(x)+\frac{x^{2/3}}{\omega(x)} \;\;&,&\;\; h(y)=\nu(y)+\frac{y^{2/3}}{\nu(y)},
\end{eqnarray}
and the continuity of $\dot{a}$ at the separation point can be written as \cite{Ramirez1}
\begin{equation}
\label{continuity}
    g(x_s)=h(y_s),
\end{equation}
where $(x_s,y_s)$ are the quantities above defined evaluated at the scale parameter of the separation surface $a_s$. Equation (\ref{continuity}) can be inverted, so we can write 
\begin{equation}
 y_s(x_s)=h^{-1}(g(x_s)) \;\;\; , \;\;\; \mu'= \frac{a_s^4}{\alpha}\left(2^{16/3} \alpha^2 P^{4/3}(a_s) y_s^{2/3}(x_s)-\beta^2\right).
\end{equation}
Since $\alpha>0$, $x$ and $y$ must always be positive. The function $y_s(x_s)$ is proven to be monotonically increasing and well-defined for all positive values of $x_s$ \footnote{Although $\nu(y)$ is non-real for $y>1/2$ it turns out that $h(y)$ is well-defined and real for all real values of $y$, since it can be written for $y>1/2$ as $h(y)=2y^{1/3} \cos(\theta(y)/3)$, where $\theta(y)$ is the argument of $1-y+\sqrt{1-2y}$ in the complex plane.}.

The splitting solution can be constructed whenever the relative acceleration of the infinitesimally separating shells has the right sign to separate them further. In \cite{Ramirez1} it is shown that this is the case when

\begin{equation}
\label{criterion}
x^{-2/3}_{\infty}\left(1-\frac{p(a_s)}{\sigma}\right)\left(1+\frac{\rho(a_s)}{\sigma}\right)^{1/3}>y_s^{-2/3}\xi(x_s),
\end{equation}
where
\begin{equation}
x_{\infty}=\lim_{a \to \infty} x(a) = \frac{\beta^3}{\kappa^4\alpha\sigma^2}  \;\;\; and \;\;\;  \xi(x_s)=\frac{\frac{2y_s^{2/3}}{y_s^{1/3}-x_s^{1/3}}-3y_s \frac{dh}{dy}(y_s)}{h(y_s)-3y_s \frac{dh}{dy}(y_s)}.
\end{equation}

In \cite{Ramirez1} a number of necessary and sufficient conditions to satisfy inequality (\ref{criterion}) have been addressed. To the purpose of this paper it is important to mention that it can indeed be satisfied for different matter-energy contents of the brane and different regions of the bulk's parameter space, and that it can only be satisfied for $\mu>0$ (which we assume hereafter).  Whenever (\ref{criterion}) holds, both the emanating false vacuum bubble solution and the original GR bulk solution, without any splitting, are possible and both comply with the junction conditions (\ref{junction}).

\section{Free energy}
\label{energy}

One way to decide among different allowed classical configurations is to compute thermodynamic potentials for each one, assuming that they represent macroscopic descriptions of more fundamental microscopic configurations whose details we can not know or simply are not of interest. It is widely believed that something of the sort applies to stationary solutions of classical gravity theories, mainly because of the existence of black hole thermodynamics and the belief in an underlying quantum gravity theory.

In this way, let us compute the difference in free energy between the two possible classical configurations we referred to in the previous section. 
%The (Helmholtz) free energy $F$ for a system in contact with a heat bath can be generically written as $\beta F  =\beta U - S$ where $\beta$ is the inverse of the temperature of the heat bath, $U$ is the internal energy and $S$ is the entropy. 
In statistical mechanics, for a system in contact with a heat bath, the (Helmholtz) free energy $F$ can be calculated by means of the canonical partition function as $\beta F = - \ln Z$, where $\beta$ is the inverse of the temperature. In the path integral approach of thermal QFT, this partition function can be approximated by 
\begin{equation}
\ln Z \approx - \hat{\cal{I}},
\end{equation}
where $\hat{\cal{I}}$ is the Euclidean action of the system valuated at the classical solution that satisfies the boundary conditions and is periodic in imaginary time with period $\beta$. Since $\mu>0$ there is a horizon in the GR-branch, so we can use the tools of black hole thermodynamics to make this calculation. We will compute the difference in $F$ by calculating the Euclidean action valuated at the competing classical solutions using the formula $\beta F = \hat{\cal{I}}$. The most important subtlety to make sense out of an action valuated at imaginary values of a time coordinate in an analytically continued spacetime is that our solutions are not stationary: although the bulk spacetime is static, the braneworld is dynamic. There are natural spacelike foliations (and hence time coordinates) for each bulk region but they do not coincide at the shell, and the angle subtended by their correspondent normal vectors at the shell varies with proper time. The absence of a natural global time coordinate makes the analytical extension gauge dependent: the properties of the complex ($\mathbb{C}\times\mathbb{R}^4$) manifold resulting from the extension of a real coordinate into the complex plane depends on the structure of the spacelike level surfaces associated with that coordinate. Nevertheless, we can formally perform this calculation as if the configuration where static and the braneworld had the properties that it has at separation time; a kind of ``adiabatic hypothesis" that we will justify by taking into account the different time scales of the system.

Let us recall the action for a manifold $M$ and its boundary
\begin{equation}
{\cal I}=\frac{1}{2\kappa^2}\int_M R-2\Lambda+\alpha L_{GB} - \frac{1}{\kappa^2} \int_{\partial M} K+2\alpha(J-2G^{ab}K_{ab}).   
\end{equation}
In order to compute $\hat{\cal{I}}$ we must calculate the on shell action for this system. After taking traces of both the field equations and the junction conditions (\ref{junction}), we have for the bulk and boundary actions the following on shell expressions
\begin{equation}
{\cal{I}}_{bulk} = -\frac{1}{\kappa^2}\int_{M} R-4\Lambda  \;\;\; , \;\;\;  
{\cal{I}}_{boundary}=\frac{2}{\kappa^2}\int_{\Sigma} K-\int_{\Sigma} S. 
\end{equation}

% INTRODUCE THE ADIABATIC APPROXIMATION APPENDIX HERE

Now we proceed to calculate the Euclidean actions for the two competing solutions at separation time. We first consider the splitting solution, its action will be labeled with the index $1$ as follows
\begin{equation}
\label{action1}
\frac{1}{2}\hat{\cal{I}}_1= \frac{2}{\kappa^2}\int_{\Sigma_v}  [K_v]-\frac{2}{\kappa^2}\int_{\Sigma_b}K_{b1}+\int_{\Sigma_b}S-\frac{1}{\kappa^2}\int_{M^+}R-4\Lambda-\frac{1}{\kappa^2}\int_{M^-}R-4\Lambda,
\end{equation}
where $\Sigma_v$ stands for the vacuum thin shell, $\Sigma_b$ for the brane, $M^+$ for the stringy branch between the thin shells and $M^-$ for the GR branch, $K_v$ denotes the trace of the extrinsic curvature of the vacuum thin shell, $K_{b1}$ denotes the trace of extrinsic curvature of the brane after the splitting (surrounded by the stringy branch), and the $1/2$ factor before $\hat{\cal{I}}_1$ comes from the $Z_2$ symmetry, as every quantity is evaluated at one of the two identical halves. 

On the other hand, the Euclidean action without the splitting, which we label with the index $0$, can be written as

\begin{equation}
\label{action0}
\frac{1}{2}\hat{\cal I}_0 =-\frac{2}{\kappa^2}\int_{\Sigma_b}K_{b0}  +\int_{\Sigma_b} S  -\frac{1}{\kappa^2}\int_{M^-} R-4\Lambda,  
\end{equation}
where $K_{b0}$ is the trace of the extrinsic curvature of the brane before the splitting (surrounded by the GR branch).

For each integral in both (\ref{action1}) and (\ref{action0}) there is a factor corresponding to the periodicity of imaginary time. For each time coordinate this period is different and the relations among them can be calculated by continuity of the metric on each thin shell. We might label these periods as $\beta^-$, $\beta^+$, $\beta^v$ and $\beta^b$, and compute $\beta^-$ by imposing regularity of the Euclidean section at the event horizon of the GR bulk, which must exist since $\mu>0$. If we evaluate $\hat{\cal{I}}_1$ {\it just after} an infinitesimal and smooth separation, the continuity of the metric implies

\begin{equation}
\label{beta}
\beta^b=(f_+(a_0)+\dot{a}_0^2)^{-1/2}f_+(a_0)\beta^+=\beta^v=(f_-(a_0)+\dot{a}_0^2)^{-1/2}f_-(a_0)\beta^-.
\end{equation}

In the same way, because we are computing $\hat{\cal{I}}_1$ as if the vacuum shell and the brane where infinitesimally separated, we might just drop the $M_+$ term in (\ref{action1}) and compute the difference in free energy between the competing configurations as follows 

\begin{equation}
F_1-F_0= \frac{1}{\beta^-}(\hat{\cal{I}}_1 - \hat{\cal{I}}_0) = \frac{4f_-(a_0)(f_-(a_0)+\dot{a}_0^2)^{-1/2} Vol(\Sigma)}{\kappa^2} \; \left([K]_{\Sigma_v}-[K]_{\Sigma_b}\right),
\end{equation}
where the first brackets denote a jump at both sides of the vacuum thin shell, while the second brackets denote the difference at the same side between having a stringy bulk and a GR bulk, and $Vol(\Sigma)$ is the volume of the braneworld at separation time.

So, the splitting would be preferred if 
\begin{equation}
  [K]_{\Sigma_v}-[K]_{\Sigma_b} = \frac{\ddot{a}_v-\ddot{a}_{\tilde{b}}}{\sqrt{\dot{a}^2+f_+(a)}} - \frac{\ddot{a}_v-\ddot{a}_b}{\sqrt{\dot{a}^2+f_-(a)}} <0, 
\end{equation}
which amounts to
\begin{equation}
\chi_1(x_s)-\chi_2(x_s) x^{-2/3}_{\infty}\left(1+\frac{\rho(a_s)}{\sigma}\right)^{1/3}\left(1-\frac{p(a_s)}{\sigma}\right) <0,   
\end{equation}
where
\begin{eqnarray}
\chi_1(x_s)&=& \frac{\frac{2}{y^{1/3}-x^{1/3}}-3y^{1/3}h'}{\sqrt{h+2y^{1/3}}}-\frac{\frac{2}{y^{1/3}-x^{1/3}}-3x^{1/3}g'}{\sqrt{g-2x^{1/3}}} , \\   \chi_2(x_s)&=&\frac{h-3yh'}{\sqrt{h+2y^{1/3}}}-\frac{g-3xg'}{\sqrt{g-2x^{1/3}}} .
\end{eqnarray}

Let us now define the function
\begin{equation}
 \Phi(x)\equiv\frac{\chi_1(x)}{\chi_2(x)}. 
\end{equation}
A numerical exploration convinced us that the function $\chi_2(x)$ is negative for all positive $x$.
% Check that chi_2 is indeed negative
%It can be shown that $\Phi(x)$ is positive, monotonically increasing and tends to a constant. 
%We are not sure how Phi(x) behaves!! 
%IT LOOKS LIKE THE SEPARATION STABILITY CONDITION IS EQUIVALENT TO THE THERMODYNAMIC STABILITY CONDITION!!!!! 
In this way, if the splitting construction can be made and it is preferred over the non-splitting solution then we should have
\begin{equation}
\label{criterion2}
    y_s^{-2/3}\xi(x_s)<x^{-2/3}_{\infty}\left(1+\frac{\rho(a_s)}{\sigma}\right)^{1/3}\left(1-\frac{p(a_s)}{\sigma}\right)<\Phi(x_s) ,
\end{equation}
where the lower bound, as analyzed in \cite{Ramirez1}, is positive, monotonically decreasing and tends to a constant. From (\ref{criterion2}) it is straightforward to find a necessary condition for having a thermodynamically preferred splitting solution: $\Phi(x_s)>y_s^{-2/3}\xi(x_s)$. Although we do not have yet a full-fledged proof, based on numerical evaluation of the functions at both sides of this inequality, there is strong evidence that $\Phi(x_s)=y_s(x_s)^{-2/3}\xi(x_s)$. If this holds then (\ref{criterion2}) can not be satisfied, which amounts to say that {\it whenever the splitting is possible, the non-splitting solution is thermodynamically preferred}.

If we take this computed thermodynamic preference at face value, it represents a possibility of spontaneous ``decay'' of a splitting solution into the non-splitting one, and not necessarily a selection rule. Although the non-splitting solution is preferred, there is a probability of spontaneously creating a false vacuum bubble with the properties described in \cite{Ramirez1}. Anyway, as described there, such a mechanism is not necessary in order to have a splitting because they are legitimate classical solutions for the given initial configuration. The thermodynamic preference is a way to assess the plausibility of the competing solutions, since the probability rate of spontaneous decay would render the splitting solution less likely than the non-splitting ones.

\section{Definiteness of the adiabatic hypothesis}
\label{definiteness}

In this work we resort to an approximation in which we compute semiclassical quantities as if the spacetime were static. The dynamical nature of the system is expressed in the time dependence of the scale parameter of the braneworld and hence the most important scale is the one set by the inverse of the Hubble parameter (the Hubble radius). The approximation makes sense as long as the timescales involved in the calculation are much shorter than the Hubble radius.
%We computed Euclidean actions by analytically extending a sort of ``locally static" spacetime. The mechanism is to extend into a neighborhood around a surface $\tau=\tau_0$ the intrinsic metric  defined there. This surface is not smooth at the brane.
%The analytically continued spacetime is likely not satisfying the junction conditions as it ignores the role of $\dot{a}$, but it constitutes a good approximation of the dynamical one as long as the time span of the extension is much shorter than the Hubble radius. On the other hand, 
There is a natural time scale in the Euclidean section: the periodicity $\beta$, chosen to have the right value in order to avoid a conical singularity at the horizon of the bulk BH. This naturally lead us to write the adiabatic hypothesis as follows

\begin{equation}
\label{adiabatic}
    H^{-1}>>\beta.
\end{equation}

As explained in the previous Section, particularly through (\ref{beta}), the periodicity in imaginary time for each region and boundary is set by the one needed in the black hole region in order to avoid a conical singularity at the horizon, $\beta^-=4\pi/f'(r_h)$. We can analyze the regime in which (\ref{adiabatic}) holds by considering
\begin{equation}
f'(r_h)=\frac{r_h}{2\alpha} \left[1-\frac{r_h^2\left(1+\frac{4}{3}\alpha\Lambda\right)}{4\alpha k +r_h^2}\right],
\end{equation}
which allow us to express $\beta^-$ as a function of $r_h$, which in turn can be obtained by solving 
\begin{equation}
-\xi A^{1/2}_{\mu}(r_h)=\frac{4\alpha k}{r_h^2}+1.
\end{equation}

 In this work we focus on GR-branch initial configurations, so we take $\xi=-1$. We will also restrict our assessment on the plausibility of the adiabatic hypothesis to the planar case ($k=0$) since its the one that most naturally leads to a standard cosmological evolution at late times. With these hypotheses the above expressions are much simpler. We can write
\begin{equation}
    \beta^-=\frac{3^{3/4}2^{3/2}\pi}{(-\Lambda)^{3/4}\mu^{1/4}},
\end{equation}
and the Hubble radius of the central brane can be expressed as follows
\begin{equation}
\label{hradius}
    H^{-1}(a_0)=\frac{2\alpha^{1/2}}{[2^{5/3}\alpha P(a_0)^{2/3}g(x(a_0))-1]^{1/2}}.
\end{equation}

We now prove that (\ref{adiabatic}) can be satisfied by choosing an appropriate value of $a_0$. One might just set $\mu$ large enough in order to ensure (\ref{adiabatic}) and then set $a_0$ also large enough to avoid having a departure from standard Friedmann evolution at late times. The Hubble radius (\ref{hradius}) depends on $x(a_0)$ and $P(a_0)$, which in turn can be written in terms of $A_{\mu}(a_0)$ and $\rho(a_0)$. From the definition (\ref{AP}) we see that $A_{\mu}(a_0)$ is a function of $\mu/a_0^4=(\mu/a_{now}^4)(z_0+1)^4$. On the other hand, the functional form of $\rho(a)$ can not depend on any chosen initial value of $a$ since, for spatially flat universes, the scale factor does not have an intrinsic geometric meaning, it can be regarded as an arbitrary initial value and all quantities valuated within the braneworld can be expressed as functions of redshift. 
%As usual in cosmological scenarios, for a combination of non-interacting linear barotropic fluids $\rho$ is the sum of terms of the form $\rho_i(z)=\rho(a_{now})(1+z)^{3(1+\omega_i)}$. 
One might just set $a_0$ such that $a_{now}=a_0(1+z_0)$ is large enough to have $\mu / a^4_{now}$ within phenomenological limits of dark radiation. As explained in Appendix B of \cite{Ramirez1}, the effect of the Friedmann evolution of the braneworld setting in the large $a$ limit is dark radiation, whose associated density parameter can be written as
\begin{equation}
\Omega_{dr}=\frac{1}{16\sqrt{1+\frac{4}{3}\alpha\Lambda}} \frac{(g(x_\infty)-D(x_{\infty}))x_{\infty}^{-1/3}}{H_0^2} \left(\frac{\mu}{a_{now}^4}\right),
\end{equation}
where $D(x)$ is
\begin{equation}
D(x)\equiv \frac{1}{(1+2x)^{1/2}}\left[\omega(x)-\frac{x^{2/3}}{\omega(x)}\right].
\end{equation}     
In this way, it is always possible to chose parameters compatible with any given realization of the standard model at late times such that the adiabatic approximation at any relevant redshift (when the splitting is possible) makes sense. Nevertheless, one might ask whether there is the same freedom in choosing suitable parameters for the $k=\pm 1$ cases \footnote{In those cases $a_0$ has a geometric meaning for the instrinsic curvature: it is proportional to the radius of curvature.} and how to evaluate thermodynamic potentials when the adiabatic approximation does not hold, but these are outside of the scope of the present paper.

\section{Final Comments}
\label{conclusions}
It is well-known that pure GR, in vacuum, admits a well-posed initial value formulation. This is also true when Einstein equations are sourced by (and hence minimally coupled to) reasonable fields such as Maxwell fields, Vlasov matter, perfect fluids or scalar fields. Well-posedness implies uniqueness of solutions for suitable initial data. In any case, the uniqueness results associated with these systems hold for initial data contained within certain functional spaces, typically Sobolev spaces $H^s\times H^{s-1}$ for data $(\gamma,K)$ on a riemannian Cauchy surface. In the case of vacuum this is known to be true for $s=2$ \cite{Klainerman} and in the case of scalar fields, Maxwell fields and certain fluids this has been proven for $s>5/2$ \cite{Parlongue}. These results do not include distributional, delta-like, initial data such as the corresponding to a thin shell. 

As mentioned in the Introduction, there are, in fact, counterexamples to the uniqueness of solutions for thin shell initial data, here we mention a few of them. For instance, the constructions made by Langlois {\it et al.} \cite{Langlois} and Gravanis and Willison \cite{GravanisWillison3} for addressing colliding thin shells or the possibility of honeycomb-like structures generated by thin shells are counterexamples within the framework of braneworld gravity. When two or more thin shells coincide at a given moment of time it is possible to construct several different outcomes after the collision. This feature is explicitly addressed in a previous paper by this author \cite{Ramirez2} in the case of a thin shell composed by several non-interacting matter fields. The non-interacting constituents can be infinitesimally separated and then one can compute the relative acceleration between them. In certain cases the constituents would separate further and represent a {\it splitting thin shell} solution, that is, a thin shell that separates into two or more different surfaces in a smooth way. For EGB gravity, the main reference of this work \cite{Ramirez1} is another example of this type of construction when the separating shells can be vacuum thin shells, as explained in Section \ref{EGBsplitting}. These references feature counterexamples to the uniqueness in the evolution of thin shell initial data because of the fact that whenever a splitting solution can be done a non-splitting solution is equally possible (a single thin shell evolving with all matter fields on it), and in both situations Darmois-Israel junction conditions, or the suitable generalization of them, are satisfied.

This work explores a way to assess this indeterminacy in those classical settings. It might represent another useful criterion to address and classify solutions in this context, and might help to determine the most reasonable outcome in situations where there is a lack of uniqueness in the evolution according to the classical theory.  Anyway, this particular assessment makes sense only if the bulk solutions feature black holes. In that case, the methods developed here can be also used in the other settings above described as long as some analogous to Eq. (\ref{adiabatic}) can be ensured. 

%These extensions are a matter for future work.  

% can Langlois et al be addressed this way? what's the indeterrminacy? there are some free parameters for the outcomes... 

In the context of the thermalon-mediated phase transitions already explored in the literature \cite{Thermalon1,Thermalon2,Thermalon3} this paper can be regarded as a part of a series of results regarding the semiclassical thermodynamic stability of thin shell solutions in EGB gravity with bulk black holes, the first one in the context of braneworld gravity. The main result, Eq. (\ref{criterion2}), suggests an equivalence between a semiclassical thermodynamic analysis and a purely classical existence condition that deserve further consideration. This equivalence might signal a formal analogy between the involving variables or be the consequence of a deeper geometrical relationship. Extensions of this work to other settings might shed light on this equivalence and are matter of future work.     

%discuss here the relationship with these and Camanho et al!!!

\section{Acknowledgments}
The author acknowledges Ernesto Eiroa, Gaston Giribet and Nicol\'as Grandi for useful conversations and insights. He also acknowledges IATE for hospitality during the early stages of this work. MAR is supported by CONICET.

%\appendix

%\section{Non-uniqueness of the evolution of thin shell initial data}
% consider to move this section to an appendix or to the introduction

%\section{On the analytic extension of a dynamic thin shell spacetime}

%As shown in Appendix \ref{definiteness}, the adiabatic hypothesis we laid out is not always applicable. In this section we try to make precise the idea of an analytic extension of an almost-everywhere static spacetime. In the situations considered the bulk geometry is static, but the brane(s) is(are) dynamic. In a SMS braneworld setting with a Lambda-vacuum bulk all the dynamics is described by the scale factor $a(\tau)$. A straightforward way to analytically extend this spacetime requires to analytically extend the $a(\tau)$ function to complex $\tau$ values. 

%As mentioned in Section \ref{energy}, there is no natural spacelike foliation for this spacetime since $t=cte$ surfaces at each bulk region do not smoothly match at the shell. This is a problem because, in general, the geometry of the imaginary sections of extensions performed using two different global time coordinates can be geometrically different, rendering the whole procedure meaningless. Let us see this with a particular example.
%Let us consider an oscillating shell made of a photon gas... 

\bibliography{biblio.bib}%

\end{document}